\documentclass[intlimits,twoside,a4paper]{article}

\usepackage[cp1251]{inputenc}
\usepackage{xcolor}
\usepackage{cmpj3}

\usepackage{bm}

\issue{2023}{26}{2}{23602}
\doinumber{10.5488/CMP.26.23602}

\title[Statistical theory of individual activity coefficients of electrolytes including multiple ionic charges]%
{Statistical theory of individual activity coefficients of electrolytes including multiple ionic charges\footnote{Dedicated to the 100th birthday of G\"unter Kelbg (1922-1988).
}}

\author[W. Ebeling, H. Krienke]{W. Ebeling\orcid{0000-0003-0740-3016}\refaddr{label1}\thanks{Corresponding author: \email{ebeling@physik.hu-berlin.de}.},
	H. Krienke\refaddr{label2}\thanks{\email{Hartmut.Krienke@chemie.uni-regensburg.de}.}}
\addresses{
	\addr{label1} Institute of Physics, Humboldt University, Berlin, Germany
	\addr{label2} Institute of Physical Chemistry, University of Regensburg, Regensburg, Germany
}

\date{Received August 11, 2022, in final form December 16, 2022}
\begin{document}
	
\maketitle

\begin{abstract}
In previous work we developed a new statistical method for calculating the individual activities of ions
including the association of ions. Here we study multi-particle
electrostatic interactions connected within higher cluster integrals and identify the ionization
constants of the mass action law of associating ion clusters. In contrast to Bjerrum and Fuoss, our concept of association is not based on spatial criteria, but instead on the strength of interaction measured in powers of the Bjerrum parameter ($e^2 / D_0 k_{\text{B}} T a$ ; $a$ is contact) and defined by asymptotic properties of the cluster integrals. For ion pair formation our mass action constant
is the classical counterpart of Planck's famous hydrogenic partition function. As a rule, the new association constants are smaller than traditional expressions, e.g., by Fuoss and Kraus, in the interesting regions of interaction parameters about fifty percent.
Several examples including CaCl$_2$, MgCl$_2$, Na$_2$SO$_4$, K$_2$SO$_4$, LaCl$_3$ and a model of seawater are studied.
For several associating electrolytes and seawater, reasonable agreement with experiments and Monte Carlo results is achieved.
\printkeywords
\end{abstract}

\section{Introduction}\label{intro}

The problem of individual ionic activities, osmotic coefficients and ion association are key problems in the theory of electrolytes \cite{Harned,Robinson,Hemmer,FaEb71,Falkenhagen,Barthel,Friedman,YukhnovskyHolovko80}.
Usual approaches to association effects in electrolytes with multiple charged ions are based on the classical concepts of Bjerrum, Fuoss and Kraus \cite{Harned}, which define pairs and triples as special
spatially defined configurations. Our concept of electrostatic association is completely different from that, we do not use spatial criteria, but the strength of interaction measured in powers of the parameter $b_0 = (e^2 / D_0 k_{\text{B}} T a)$ ($a$ is contact distance of ions). We follow the concepts of Onsager, who stated in 1968 at a conference in Montpellier: ``Bjerrums choice is good but we could vary it within reason. In a complete theory
this would not matter; what we remove from one side of the ledger would be
entered elsewhere with the same effect'' \cite{Hemmer}.
In our concept for the definition of pairs we use this freedom and assume that pair-associates are formed by higher order (negative) contributions of binary charge interactions $b_0^n$ with $n \geqslant 4$
to the pressure and other thermodynamic functions. Triple and quadruple association is generated by (negative) contributions of three or four charges with opposite signs contributing higher orders in $b_0$ to the pressure. Such a definition of association may seem less transparent in comparison with spatial definitions, although it is  easier in the light of statistical thermodynamics. The background of the present work
are the cluster expansions for systems with Coulombic interactions based on the fundamental papers of Joseph Mayer since 1949 which were developed in the 50th and 60th by E. Haga, E. Meeron, H. Falkenhagen, G. Kelbg, I. R. Yukhnovskii, H. Friedman and others.
Note that the cluster expansions were developed in two versions, based on density and those based on fugacity expansion~\cite{KeEb70, KeEb70-1}.
Both versions are connected with diagrammatic expansions with respect to the interaction strength $e^2$ or in the case of hard charged spheres with the Bjerrum parameter $b_0 = e^2/D_0 k_{\text{B}} T a$.
An important role played personal meetings in Rostock and Lviv between H. Falkenhagen, G. Kelbg, I. R. Yukhnovskii, H. Friedman
and the present authors who witnessed these discussions. 

The basic concepts which we use here were presented first in \cite{FaEb71,Falkenhagen,EbKe66,KeEbKr68,Eb67,Kelbg72,FriedmanEb79}
and were  developed later in \cite{Barthel,KrBaHo00,EbHiKr02,EbFeKrRG19}.
These concepts are  mainly based on a mathematical analysis of the cluster functions of the Mayer-theory of ionic solutions \cite{FaEb71,Falkenhagen,Friedman}. This basic theory of ionic solutions led to cluster expansions, which were rederived and generalized on the basis of collective variables by Yukhnovskii \cite{YukhnovskyHolovko80} and Kelbg \cite{Kelbg72}.
The connections between cluster expansions and association theory were established in \cite{FaEb71,EbKe66,KeEbKr68}. The main idea
of this concept is that the contributions of ion pairs and ion triples
are given by certain relevant parts of the second and third cluster integrals in the pressure expansion. 
The parts relevant to bound states
are in the low temperature asymptotically large and negatively definite,  this way strongly decreasing the pressure.
These contributions are of higher order in the interaction parameter $e^{n}$, e.g., $n \geqslant 8$ for pairs. These concepts lead to quite transparent definitions of the ionization constants for pairs and triples as we showed in the foregoing work \cite{FaEb71,Falkenhagen,EbFeKrRG19,EbGrRG22}.
Alternative concepts of electrostatic association were considered in many works
\cite{FaEb71,ValiBoda,EbGr80,EbGr80OL,Justice, Justice1, Justice2, KrBa98,Schroer11}.

\begin{figure}[htbp]
\begin{center}
\includegraphics[height=5cm,width = 5cm,angle=0]{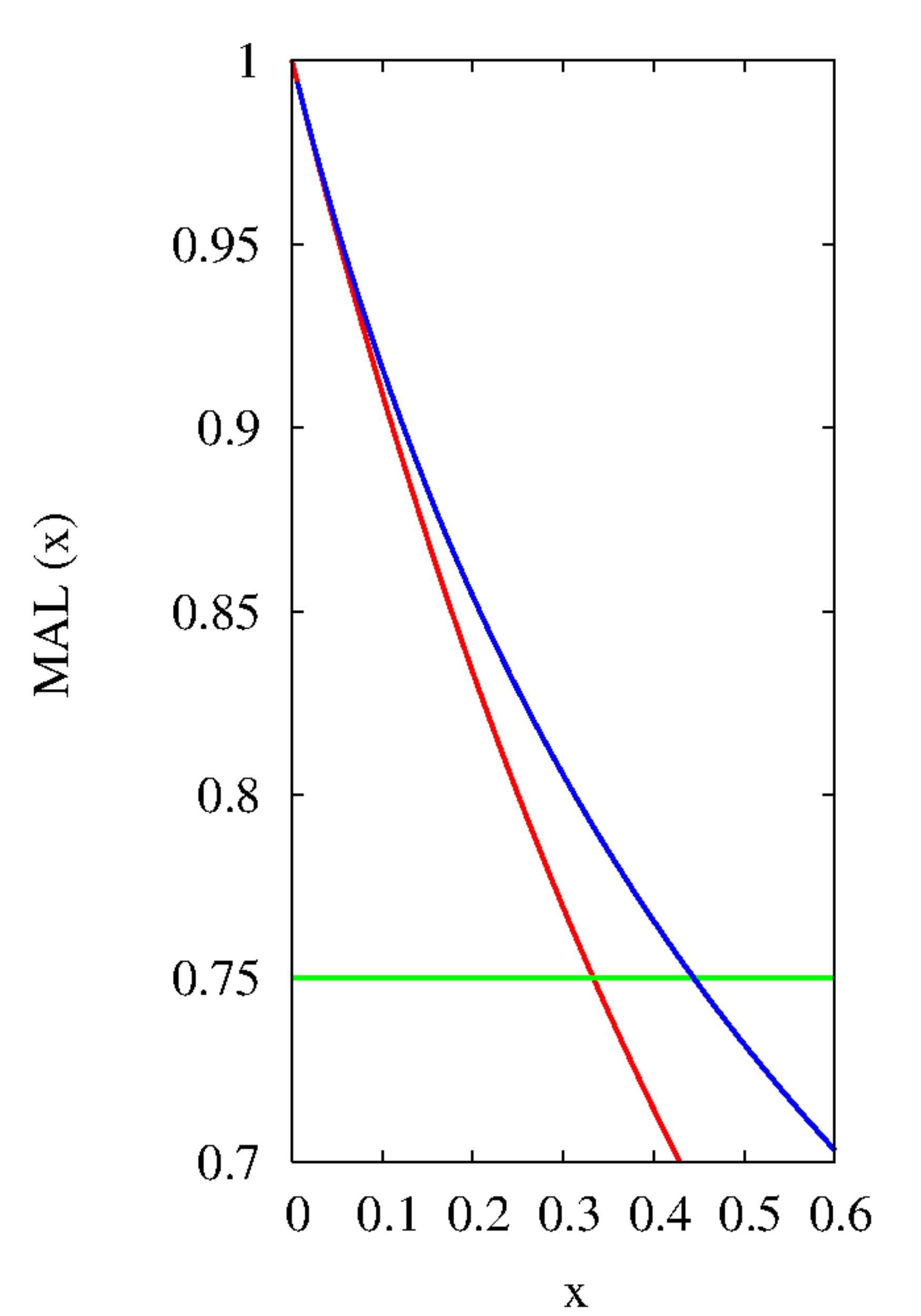}
\includegraphics[height=5cm,width = 5cm,angle=0]{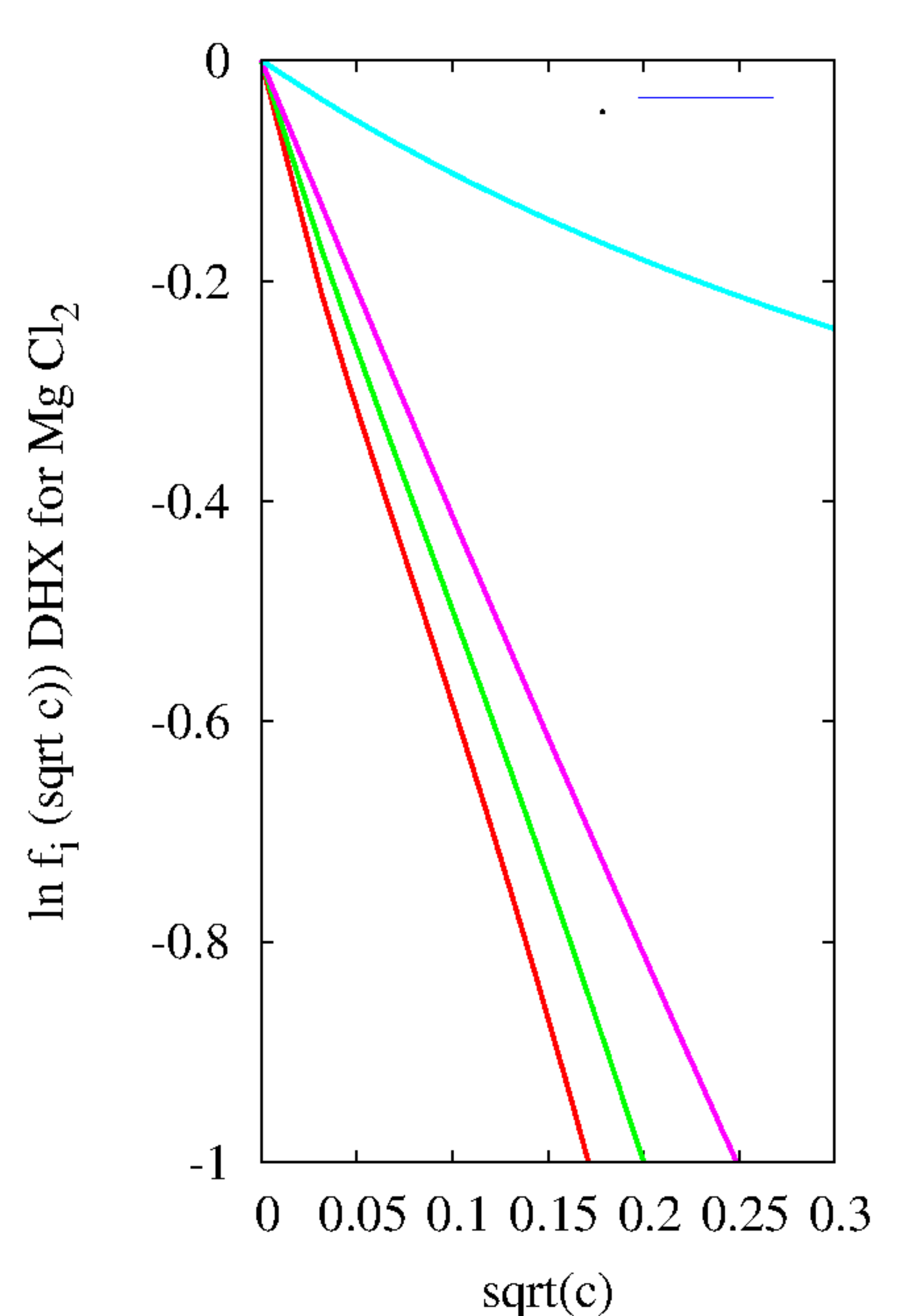}
\caption{(Colour online) Left-hand panel: The range of validity of our semi-chemical, i.e., using a first order rational polynomial approximation, (red curve) in comparison to the full nonlinear mass action law
for ion pairing (blue curve). For a simple model, our approach stops working if more than
about 25 percent of the ions are associated i.e., $\alpha_i < 75$.
Right-hand panel: The activity coefficients for the solution of MgCl$_2$ calculated within the present approach. We show from below the activity coefficients of Mg$^{2+}$, then the mean activity of 
MgCl$_2$ and then
 the activity of the ion Cl$^-$. The upper curve (turquoise) shows the log of the degree of ionization
 $\ln (\alpha)$.
}
\label{Valleau}
\end{center}
\end{figure}

The influence of electrostatic pair and triple association of ions is of high relevance for many problems
where electrolytes with higher charges such as MgSO$_4$, MgCl$_2$ or Na$_2$SO$_4$ play a role, e.g., in
seawater research~\cite{Harned}.
In recent works \cite{EbFeCa20,EbFeKrRG19,EbGrRG22}, the present authors summarized the existing
theoretical knowledge on the activity coefficients and the individual conductivities of electrolytes in such systems. We responded this way to urgent
needs for extending and deepening the knowledge on the physico-chemical properties of the
components of seawater and other complex natural and technological electrolytes
which include higher charges. For example in seawater in addition to univalent ions, such as Na$^+$, K$^+$, Cl$^-$ also the double-charged ions
Mg$^{2+}$, Ca$^{2+}$ and SO$_4^{2-}$ are of importance for seawater properties.
In many technologies several multiple-charged ions like the vanadium ions are becoming of substantial interest, e.g., for modern battery development.
Here, we develop a new look at the analytical theory of ions with higher charges, where the differences between the individual and the mean activities are large \cite{Harned,Robinson,Barthel,EbFeKrRG19,Sakaida11}. We devote special attention
to the consequences of charge asymmetry and higher charges to the individual ionic activities.
In previous work, triple association was mainly neglected, while here we give a more systematic approach.
In our statistical approach we use the ion densities $n_i$ (in particle numbers per cm$^3$) or the molarities $c_i$ (in moles of ions per dm$^3=$ liter) as the basic primary quantities. For seawater we  also use the
chlorine molality ($m_{\text{Cl}}$, in moles of chlorine per mass of pure water), and the salinity
($S_A$ is the mass of dissolved sea salt per mass of seawater). Using the $n$-scale, the activity coefficients are denoted by $f_i$ and the so-called practical (or molal) activity coefficients $\gamma_i$ are defined using molalities, $m_i$ \cite{Barthel,Friedman}
where the molalities $m_i$ are measured in moles per mass of solvent \cite{Friedman}.
Further we use the partial osmotic coefficients $g_i$ derived
from the osmotic pressures $P_i$ of species $i$. We use standard methods for estimating  individual activities and osmotic coefficients
as the virial expansion of the thermodynamic functions \cite{FaEb71,Falkenhagen,Barthel,Friedman,EbFeKrRG19,EbeSch1983,Pitzer,Simonin}.
The approach given here is restricted to ion association problems, where less than about $1/4$ of the ions are in associated states. This is due to our quasi-linear approach to the mass action law and to the first order approach to rational pressure expansions.
The idea behind our approach is that a complete physical description of any higher interaction orders includes everything
and in particular the chemical effects. The problem is, however, to identify the responsible higher order terms and work out
these contributions relevant for association. We have described this here in some detail for pairing effects
and in less detail for triple and quadruple association effects. A restriction of our method is that the nonlinearities contained already in a simple mass action law, are included in our procedure only in a quasi-linear way.
As shown in figure~\ref{Valleau}  for a simple model, the results for the degree of ionization of individual ions using our simple method which is
half way between physical
and chemical description, is for weak association in reasonable agreement with the complete mass action law.
The price to pay for the simplicity of our method is that we have to stay in the region of weak association effects,
say that the degrees of ionization are higher than 75 percent.

\newpage
\section{Interaction potential and association functions }

We follow here basically the concepts developed in \cite{EbFeKrRG19}. First we define the potential of the mean force between the ions $i$ and $j$ in the solutions as $\psi_{ij}$. The average potentials and forces are split into a Coulombic and a short-range part
\begin{align}
 \psi_{ij} = V_{ij}(r) + V_{ij}'(r).
\end{align}
The electrical part is given by Coulomb's law
\begin{align}
V_{ab}(r) =  k_{\text{B}} T \cdot \frac{\ell}{r}; \quad \ell_{ij} = Z_i Z_j \ell, \quad \ell = \frac{e^2}{D_0 k_{\text{B}} T}; \quad D_0 = 4 \piup \epsilon_0 \epsilon_r,
\end{align}
where $\epsilon_r (T, p)$ is the relative dielectric constant of pure water and $\ell$ the Coulomb length (also called Landau length or with the pre-factor $1/2$ the Bjerrum length). Both are functions of
temperature and pressure.

In what follows we perform all calculations for the temperature $ T = 298.15$ K (i.e., 25 degrees Celsius).
and assume for the relative dielectric constant the value $\epsilon_r = 78.36$. Then, we get for the Coulomb length $\ell = 715.4$ pm.
In our model this is the only parameter which is temperature  and pressure  dependent, so the transition to other values
of $p, T$ is reduced to changing $\ell$.
The short range forces are of hard-core type, where $R_{ij}$ are the contact distances. These are the most important key data
in our approach.
Several values for the contact distance are given in table~\ref{TabContact}. Most of the values were given already in~\cite{EbFeKrRG19}. We are of opinion that the most important data about contact distances may be obtained
from MC and MD simulations
as presented, e.g., in \cite{Krienke13,ValiBoda}.
We have added a few not so well studied ions as La and Cd. The crystallographic radius for
Cd$^{2+}$ is with 95 pm just a few pm higher than that for Mg$^{2+}$ which is 86 pm. Therefore, we may assume that the contact distances in solution are also close, we took
$R_{\text{Cd} \text{Cl}} = 420$ pm. For La$^{3+}$, we know that the crystallographic radius is smaller than that for Mg$^{2+}$. Following canonical MC
simulations by Valisko and Boda we
assume for those ions in water $R_{\text{LaLa}} = 430$ pm and $R_{\text{LaCl}} = 270$~pm corresponding to a quite large Bjerrum parameter
$\xi_{\text{pm}} = 7.95$.
The pair association constants given in table\ref{TabContact} stem from an earlier theory, the triple association constants from new calculations.

The definition of a mass action constant introduced above applies to all charge-symmetrical ionic associates
including, e.g.,
Mg$^{2+}$-SO$_4^{2-}$. More
difficult is the question how to define the association constant for triple associates as $(+)(--)(+)$ and $(-)(++)(-)$.
We search first for the maximum binding energy of 3 ions, e.g., Cl-Mg-Cl or Na-SO$_4$-Na one of them double charged. The biggest energy has the
linear configuration $(-) (2+) (-)$ or $(+) (2-) (+)$ in the linear arrangement of the ions in contact. A simple estimate
of the energy of 3 ions in contact gives us
\begin{align}
E_{-(2+) -} = E_{+ (2-) +} = (7/2) E_{+-}\,.
\end{align}
Consequently, we expect the asymptotic
\begin{align}
k_{-(2+)-} \sim \exp[3.5 |E_{+-}| / k_{\text{B}} T].
\end{align}
This result corresponds to the estimates proposed by
Friedman and one of these authors \cite{FriedmanEb79} by using the results of mathematical studies
of cluster integrals \cite{Friedman,EbKe66}. Including the pre-factors, this estimate of the asymptotic reads
\cite{Friedman,EbKe66,FriedmanEb79}
\begin{align}
k_{aba} \simeq 8 \piup^2 R_{ab}^6 \cdot \frac{\exp[(2 \ell_{ab} / R_{ab})]}{\left[(\ell_{ab} + \ell_{aa}) / 2 R_{ab}\right]} \cdot \frac{\exp[(\ell_{aa} / 2 R_{ab})]}{(- \ell_{aa} / 2 R_{ab}) }.
\end{align}
This estimate was obtained in Kelbg's early work \cite{EbKe66} by using the assumption that the integrands have a cusp at contact, which provides the asymptotic.
A qualitative  procedure to treat triple association of the type $(- + - )$ or $(+ - +)$ is the method of effective charges as known from the
treatment of helium formation as bound state of two electrons and one two times positively charged alpha-particle.
The bound state energy of an electron-alpha pair is $Z^2 E_H$ where $Z =2$ is the charge of the He-nucleus and the corresponding
pair association constant is
\begin{align}
k_2 \sim \exp[Z^2 E_H].
\label{HydrogenMA}
\end{align}
In one of his latest but quite important works, Max Planck showed that for Coulombic systems,
here for hydrogen-like bound states, the exponential function should be replaced by a cropped exponential, where the first two terms of the exponential are cropped
\begin{align}
\exp[Z^2 E_H] \rightarrow \exp[Z^2 E_H] - 1 - Z^2 E_H .
\label{Hydrogencrop}
\end{align}
Plancks partition function begins with the order $e^8$ similar to the Falkenhagen partition function \cite{FaEb71}.
For a pair of single charged ion and a Z-fold charged ion, the Falkenhagen theory provides with $b_0 = \ell/R_{+-}$
the expression \cite{FaEb71}
\begin{align}
k_2 = 8 \piup R_{+-}^3 m (Z b_0) ; \qquad b_0 = \frac{\ell}{R_{+-}}; \qquad m(x) = \sum_{m=2}^{\infty} \frac{x^{2m}}{(2m)! (2m-3)}.
\label{eq8}
\end{align}
The arguments which lead to this series on the Bjerrum parameter are completely different from Plancks arguments.
Therefore, we consider the relations between the quantum theory of Planck and the classical approaches
by Bjerrum and Falkenhagen \cite{FaEb71} more in  detail.
The fundamental work of Max Planck published in 1924 is notoriously difficult to understand, which is the reason why this work is not so often cited as the earlier papers by Planck. In order to apply Planck's ideas to classical systems, we come back to our own paper, which is an exercise about Planck's work with more detailed calculations \cite{Eb67}.
The question to be studied in detail is, how  the definition of the Falkenhagen
partition function equation~(\ref{eq8}) is related to the Planck concept equation (\ref{Hydrogencrop}).
In Planck's quantum statistics, the basic assumption is
\begin{align}
k_2 = 4 \piup \int_0^{\infty} \rd r r^2 S_{+-}(r;b) ,
\label{eq9}
\end{align}
where $S_{+-}(r;b)$ is the bound state part of the quantum pair probability. This term is proportional to the sum of probabilities to find an opposite charge in a bound state around the central charge, which is expressed here by the diagonal part of the density matrix. In order to find the classical counterpart of this function first we find  the probability density that a pair ($+-$) is in distance $r$ in states with negative relative energy
\begin{align}
S_{+-} (r;\epsilon < 0) = \frac{4 \piup}{(2 \piup \mu k_{\text{B}} T)^{3/2}} \int_0^{p_0} \exp[- \beta p^2 / 2 \mu + g (r)],
\qquad g(r) =  \ell_{+-} / r .
\end{align}
Here, $p_0 = \sqrt{2 \mu Z e^2 / \epsilon r}$ is the momentum where the pair energies change the sign from negative to positive values ($\mu$ is relative mass). Carrying out the integral over all momenta $p < p_0$, first we get  an integral over the error functions and then explicitly the following series \cite{Eb67}
\begin{align}
S_{+-}(r;\epsilon < 0)) = \exp[g(r)] \phi(\sqrt{g(r)}) - (2 / \sqrt{\piup}) \sqrt{g(r)} =
\frac{2}{\sqrt{\piup}} \sum_{s = 1}\Big[\frac{2^s g(r)^{s + 1/2}}{(2 s +1)!!}     \Big].
\end{align}
The integral over the distances of this complicated function is divergent, so we have to omit according to Plancks arguments the first two terms which correspond to states close to $\epsilon = 0$ and treat them together with the free states. Since $g(r) \sim \ell$, these first terms correspond to low orders in the interaction $e^2$. The remaining higher order terms $g(r)^{s+1/2}$ with $s > 2$ provide us with the wanted bound state contribution
\begin{align}
S_{+-}(r;b)) = \frac{2}{\sqrt{\piup}} \sum_{s = 3}\Big[\frac{2^s g(r)^{s + 1/2}}{(2 s +1)!!} \Big].
\end{align}
The integral over $r^2 \rd r$ is now convergent and gives a finite integral which is the direct classical counterpart of Plancks equation (\ref{Hydrogencrop}).
Investigating this integral over the radial coordinate in detail, we find out that it has a special property \cite{Eb67}
\begin{align}
 4\piup \int_{0}^{\infty} \rd r r^2 S_{+-}(r;b) = 2 \piup \int_{0}^{\infty} \rd r r^2 [\exp(g) + \exp(-g) - 2 - g^2/2] .
\end{align}
We may conclude for the same integral with the lower limit $R_{+-}$:
\begin{align}
k_2 &= 4\piup \int_{R_{+-}}^{\infty} \rd r r^2 S_{+-}(r;b) =  4 \piup \int_{0}^{\infty} \rd r r^2 S_{+-}(r;b) \nonumber \\
 &= 2 \piup \int_{0}^{\infty} \rd r r^2 [\exp(g) + \exp(-g) - 2 - g^2/2] = 8 \piup R_{+-}^3 m (Z b_0).
\end{align}
From the point of physics, this result is remarkable, since it  says that Falkenhagen's mass action constant which originally was based on more mathematical arguments than simplicity, is indeed a complete analog of Planck's famous hydrogenic mass action constant. Further this result means that equation~(\ref{eq9}) is valid in the classical as well as in the quantum-statistical case, i.e., the approach is equivalent for the classical as well
as for quantum Coulombic pairs.

For the formation of He-like triples, the mass action constant  in zeroth order, according to the idea of effective charge, can be approximated by a product
\begin{align}
k_3 \sim \exp[2 {\tilde Z}^2 E_H] \sim [k_{+-} ({\tilde Z})]^2,
\end{align}
where ${\tilde Z} \simeq 1.8$. The decrease of the charge from 2 to 1.8 is due to the screening of an interacting charge pair
by the third charge in the triple.
Here we  transfer  this effective charge approach from the helium theory to the 2--1-association problem, assuming an effective charge of the double-charged ions of about ${\tilde Z} \simeq 1.8$.
The method of effective charge also works  for electrolytes and provides a simple semi-quantitative approach to triple ionization, which provides a bit more flexibility and is in reasonable agreement with the other approaches.

We start here our investigation from the free energy in a physical description of the ions by the canoni\-cal ensemble \cite{EbKe66,EbFeKrRG19}.
General expressions from statistical thermodynamics for the cluster contributions $S_{i}^{(k)}$ to the negative
free excess energy of hard charged spheres read \cite{Falkenhagen,Friedman,YukhnovskyHolovko80,EbKe66,KeEbKr68}
\begin{align}
F_{\text{ex}} = F_{\text{DH}} + F_2 + F_3 + ...= - k_{\text{B}} T V \Big[\frac{\kappa^3}{12 \piup} \sum_{ij} \zeta_i^0 \zeta_j^0 R(\kappa R_{ij}) + \sum_{i,k} S_{i}^{(k)} \Big],
\end{align}
where $R(x)$ is the so-called ring function in Debye-H\"uckel approximation and $\zeta_i^0$ the Onsager relative screening factors defined by
\begin{align}
R(x) = 1 - \frac{3}{4} x + \frac{3}{5} x^2 - \ldots; \qquad \zeta_i^0 = \frac{n_i e_i^2}{\sum_j n_j e_j^2}.
\end{align}
The corresponding chemical potential is
\begin{align}
\mu_i = - \frac{Z_i^2 \ell \kappa} {2} \sum_j \zeta_j^0 G_0(\kappa R_{ij}) + \ldots; \qquad G_0(x) = \frac{1}{1 + x}.
\end{align}
The sums are to be extended over the species of ions $i$ and all orders of clusters $k$.
The strong coupling contributions of ions $i$ read in the cluster order $k = 2,3,4, ...$ \cite{Falkenhagen,Friedman,YukhnovskyHolovko80,EbKe66,KeEbKr68,Schroer11}
\begin{align}
S_i^{(2)} = \frac{1}{2} n_i \sum_j n_j \int \rd {\bf r}_j \left[ \psi_{ij} - \frac{1}{2} G_{ij}^2\right] , 
\end{align}
\begin{align}
S_i^{(3)} =  \frac{1}{2 \cdot 3} n_i \sum_{jk} n_j n_k \int \rd {\bf r}_j \rd {\bf r}_k \Big[\psi_{ij} \psi_{ik} \psi_{jk} +  g_{ij} \psi_{ik}\psi_{jk} + g_{ik} \psi_{jk}\psi_{ij} + g_{jk} \psi_{ij}\psi_{ik}\Big],
\end{align}
\begin{align}
S_i^{(4)} =  \frac{1}{2 \cdot 3 \cdot 4} n_i \sum_{jkl} n_j n_k n_l \int \rd {\bf r}_j \rd {\bf r}_k \rd {\bf r}_l
\Big[\psi_{ij} \psi_{jk} \psi_{kl} \psi_{il} + \ldots\Big].
\end{align}
Here the strong coupling function $\psi_{ij}$ which is of higher order $O(e^4)$ in the interactions is defined by
\begin{align}
\psi_{ij} = \exp[G_{ij} - \beta V'_{ij}] -1 - G_{ij}; \qquad
G_{ij}(r) = - Z_i Z_j \ell \frac{\exp(-\kappa R_{ij} - \kappa r)}{(1 + \kappa R_{ij})}.
\end{align}

We use here the so-called nonlinear Debye-H\"uckel approximation
for charged hard spheres proposed by Onsager and Fuoss and developed by these authors \cite{Hemmer,EbFeKrRG19,Bich}.

The contributions of the cluster theory given here including a nonlinear DH-screening contain the full 2nd virial coefficient in good agreement with available HNC data. Including the  third virial coefficient for associating electrolytes, e.g., for  MgSO$_4$ with $R = 420$ pm brings the theory even to a quite good agreement with MC data up
to nearly 1 mol/liter \cite{EbFeKrRG19}. An earlier comparison of our approximation with MC calculations \cite{KrienkeSt84} also gives a rather good agreement. The present approximation gives the values for the free energy
which are close to MC results up to about $c \simeq 1$~mol/liter  \cite{Valleau}.
This way we see that the contributions provided by the 2nd and 3rd virial coefficients are quite relevant for describing two- and three-particle association effects.
The idea behind our physical theory of association is as follows: we connect the definition of the associates with the convergent, and in the asymptotic for strong interactions, dominant parts of the cluster integrals. Correspondingly, we define the degree of ions of kind $i$ bound in pairs or triples as the asymptotic convergent strong coupling parts of the cluster integrals which lead to big negative contributions to the free energy:
\begin{align}
\delta_i^{(2)} = \sum_j n_j k_{i,j}^{(2)} (T); \qquad k_{i,j}^{(2)} (T) \sim \text{asy} \Big[\int_{\text{conv}} \rd 2 S_i^{(2)}
 \Big], \nonumber \\
 \delta_i^{(3)} = \sum_{jk} n_j n_k k_{i,jk}^{(3)} (T) ; \qquad
 k_{i,jk}^{(3)} \sim \text{asy} \Big[\int_{\text{conv}} \rd 2 \rd 3 S_i^{(3)}  \Big].
\end{align}
Asymptotic and convergent means here that the given integrals should be treated as follows: the integrals should be extended over relative coordinates and performed after identification of the
convergent and (at strong association) dominating positive definite parts of the cluster integrals.
The results of these operations depend only on the temperatures
and provide the corresponding association constants.
We underline that this is the central point of our concept of association. The problem is that the asymptotic part is not uniquely defined.
We have some asymptotic freedom. However, at this point we may use the freedom which gives us Onsagers statement, that in the ledger of the theory
only the observable sums of mass action effects and long range interaction effects are fixed and controllable by experiments \cite{Schroer11}.
We note that a systematic approach to the derivation of mass action constants from cluster integrals was developed already in \cite{Eb74} on the basis of the grand canonical ensemble.

Here, we will explain our concept on examples. An essential part of the concept is that for Coulombic association,
the classical
exponential factors,  appearing, e.g., in equation~(\ref{HydrogenMA}) should be replaced by some kind of cropped exponential functions
which do not contain the first two terms. This will be demonstrated now. For the case of
pair formation, we get the estimate of Falkenhagen given already above
\cite{FaEb71,Falkenhagen,Barthel}
\begin{align}
k_{ij}^{(2)} (T)  = 8 \piup \ell_{ij}^3 n_2 (\xi_{ij}) = 8 \piup R_{+-}^3\cdot  m(\xi_{+-}),\\
m(\xi) =\frac{1}{2}\cdot \left(K_{04}(\xi)+K_{04}(-\xi)\right) = \sum_{m=2}^{\infty} \frac{\xi^{2m}}{(2m)! (2m-3)}.
\end{align}
Here, the pair association function $m(x)$ is related to the $Ei(x)$ functions and to the so-called Kirkwood function \cite{FaEb71,Falkenhagen}.
More difficult is the question how to define the association constant for triple associates as $(+)(--)(+)$ and $(-)(++)(-)$.
We follow mathematical approaches to the pair and triple cluster integrals~\cite{FriedmanEb79,KrienkeSt84}.
The idea behind the so-called constant factor approach is as follows:
one or more factors $\psi_{ij}(r_{ij})$ in the integrands of the relevant clusters belong on the case of associating clusters to repulsive interactions.
For these factors $\psi_{ij}$, the exponent is positive expressing the repulsion of two equally charged ions.
The essential point for the integration is that repulsive ions on average cannot come closer than
certain minimal distance $2 R_{+-}$ for triples. In any case, these factors expressing repulsion are slowly changing.
Fixing the values of the repulsing functions at the most probable values, we may take the factors out of the integral.
For triple clusters, we find, e.g., assuming a double charged ion $a$ at 1 and and two single charged ions $b$ at 2 and 3, the estimate
\begin{align}
S_{\text{aba}}(1,2,3) \sim  S_{aa}(2,3) S_{ab} (1,2) S_{ab} (1,3) ,
\end{align}
\begin{align}
k_{\text{aba}} \simeq  8 \piup^2 R_{+-}^6 \cdot \left[\exp(-\ell_{aa} / 2 R_{ab}) - 1 +  (\ell_{aa} /(2 R_{ab}))\right] \cdot \left[m(\xi_{ab})\right]^2
\label{eq27}.
\end{align}
We consider the association constant of a triple consisting of
two univalent ions $j,k$ with a multiple charged ion $i$ imbedded in between.
The repulsive factor in the cluster integral for 3 ions  may be approximated by the first term in the Taylor expansion
 $(\ell_{--} / 2 R_{+-})^2 $ which including a steric $r^2$ provides just a constant factor.
Further,  in order to have more flexibility, we may introduce an effective charge ${\tilde Z}$ in agreement with previous studies of the properties of the triple cluster integrals \cite{EbKe66,KeEbKr68,KrienkeSt84}. For 2-1 associates, we have ${\tilde Z}^2 \simeq 3.5;$  ${\tilde Z} \simeq 1.8$.
We see that our definition of mass action constants does not contain contributions
in lower orders of $e^2$, the pair association constant starts with $e^8$.
The contributions of lower orders are missing since otherwise, we would have conflicts between contributions from the mass action terms with the screening contributions, in particular with the extended Debye limiting law~\cite{EbHiKr02}.
The degree of free ions of kind $i$ which is that part of the ions which are not associated in pairs, triples, quadruples etc.,
is defined as the relation between the number of free ions to the total number of those ions
\begin{align}
\alpha_i = \frac{N_i^{\text{free}}}{N_i^{\text{free}} + N_i^{\text{asso}}} = \frac{1}{1 + \delta_i^{(2)} + \delta_i^{(3)} +  \ldots}.
\end{align}
This is the typical mathematical structure we have in the semi-chemical approach suggested first
by Justice et al. \cite{Justice,Justice1, Justice2}.
The $\delta_i^{(3)}$ and $\delta_i^{(4)}$ we find in our approach as the asymptotically big negative definite contributions from the 3rd and 4th virial coefficients
providing  relevant contributions to triple and quadruple association. According to our estimates in the previous work, the pair contribution is of order $\xi^4$, the triple contribution is of order $\xi^{10}$ \cite{EbGrRG22}.
The corresponding thermodynamical functions as, e.g., the activities have, as a rule, a typical structure of rational functions which  stems from the
nonlinear Debye-H\"uckel and MSA theories. One way to arrive at these structures is an extension of the first corrections to the limiting law for the activities which reads \cite{EbFeKrRG19}
\begin{align}
\ln f_i^{el} = \ln \alpha_i -  \frac{Z_i^2 \ell \kappa}{2 } \cdot \sum_j \zeta_j^0 \frac{1}{1 + \kappa R_{ij}}, \nonumber \\ 
\quad \zeta_j^0 = \kappa_j^2 / \kappa^2, \quad \kappa_j^2 = 4 \piup \ell n_j z_i^2 ,\quad \kappa^2  = \sum_j \kappa_j^2.
\end{align}
As already mentioned, there is a close relation to an approach developed by Justice et al. \cite{Justice,Justice1, Justice2}.
Note that instead of the relative screening factor $\zeta_i^0$, used already by Onsager (however, denoted by the letter $\mu_i$), we worked in some earlier papers \cite{EbFeCa20} with the
half quantity $\zeta_j = \zeta_i^0 / 2$. For a binary electrolyte, the calculation of the $\zeta_1,\zeta_2$ is particularly simple
\begin{align}
\zeta_1^0 = \frac{n_1 e_1^2}{n_1 e_1^2 + n_2 e_2^2} = \frac{|z_1|}{|z_1| + |z_2|}, \qquad \zeta_2^0 = \frac{|z_2|}{|z_1| + |z_2|}.
\end{align}
We use here Onsagers notation, but changing his
letter $\mu_i$ which we use for the chemical potential. For for some examples of binary electrolyte, the calculation of the $\zeta_1,\zeta_2$ give, e.g., for MgCl$_2$, CdCl$_2$ the pair $(1/2, 2/3)$ and for LaCl$_3$ the pair $(1/4, 3/4)$.

We use here, in the simplest approximation for the zeroth order, the Debye-H\"uckel-approximation; more advanced
possibilities are the Mean Spherical Approximation (MSA)
and the related Henderson-Smith approximation (HSA).
The nonlinear Debye-H\"uckel-approximations (DHA) take into account the first- and second-order terms
$G_1(x), \, G_2 (x),\, {\tilde G}_1(x), {\tilde G}_2 (x)$ \cite{EbFeCa20}.
We will not explain the extension of this theory to the MSA approximation in detail and refer to other works
\cite{Simonin,Triolo,HendersonSmith,Blum80,Vilarino}. Our favorite approach is
based on a generalization of the Henderson-Smith formula for the pair distribution to  arbitrary contact distances \cite{EbFeKrRG19}.
According to this approach, the transition from the DHA to the MSA consists formally in replacing the Debye-H\"uckel parameter $\eta$ by
\cite{EbFeKrRG19}:
\begin{align}
\eta_{ij} = \kappa R_{ij} \rightarrow \gamma_{ij} = \frac{1}{2}\cdot \left(\kappa R_{ij}+ \sqrt{1 + 2 \kappa R_{ij}} - 1\right).
\label{neweta}
\end{align}
Note that the more advanced expression $\gamma$ includes higher orders in $\kappa$ and that this way programs written for DHA may be easily 
extended to the Henderson version of the MSA.
The central point in our theory is taking into account the differences of the diameters of the ions,
since the differences between the individual activities and conductivities depend strongly on
the differences between the contact distances~\cite{EbFeKrRG19,EbFeCa20}. The degrees of ionization as well as the conductivities depend mainly
on the contact distance of oppositely charged ions.

\section{Association theory using rational pressure expansions}

In the second section we identified the association contributions as the big positively definite terms in the virial expansions leading to negative contributions to the free energy.
In an alternative approach based on virial expansions of the osmotic pressure, we
construct now rational representations of the pressure, where the negatively definite
terms in pressure expansions are moved to the denominator of rational expressions. We do not need here explicit definitions
of mass action constants.
For the partial osmotic pressure contributed by the ionic species $i$ we get, by introducing the distribution function
into the virial formula,
\begin{align}
P_i = P_i^{id} + \frac{1}{3} u_i^{el} + P_i^{hc} ; \qquad P_i^{hc} = \frac{2 \piup}{3} n_i \cdot \sum_{j}  n_j R_{ij}^3 \cdot\exp \left[\frac{\xi_{ij}}{(1 + \kappa R_{ij})} \right].
\label{pDHX}
\end{align}
The electrical energy density can be expressed by Euler functions \cite{EbFeCa20,EbFeKrRG19}. We define weakly and moderately coupling, separated from strongly coupling and hard core parts of the osmotic pressure
\begin{align}
 P_i / (n_i k_{\text{B}} T) = g_i = 1 - \frac{1}{6} z_i^2 \kappa \ell \cdot \sum_j \zeta_j^0 \cdot [{\tilde G}_0 + {\tilde G}_1] (\eta_{ij}, \xi_{ij}) + g_i^{\text{sc}} + g_i^{hc}.
\label{partialP}
\end{align}
The contribution of weak coupling $G_0$ is given by a Debye-like osmotic function and $G_1$ is a correction for asymmetric ionic systems \cite{EbFeKrRG19}.
\begin{align}
{\tilde G}_0 (x) = \frac{1}{1 + x} \Big[1 - \frac{x}{2 (1 + x)} \Big]; \quad g_i^{\text{sc}} = g_i^{\text{sc}2} + g_i^{\text{sc}3} + g^{\text{sc}4}.
\end{align}
The G-functions represent the different orders in $\eta$ and
$\xi_{ij}^k$ \cite{EbFeKrRG19}. The first correction ${\tilde G}_1$ is relevant for asymmetric ionic systems
and leads to logarithmic terms
\begin{align}
{\tilde G}_1 (\xi_{ij},\eta_{ij},\zeta_i) =  \frac{\xi_{ij}}{18 (1 + \eta_{ij})^3}\cdot \left[e_1(2 \eta_{ij}) -
3 \eta_{ij} \zeta_{i} \exp(- 3 \eta_{ij})\right].
\end{align}
The strong coupling contributions to the pressure $g_i^{\text{sc}} = g_i^{\text{sc}2} + g_i^{\text{sc}3} + g^{\text{sc}4} + ...$ which are all strictly negative, stem from the second, third, forth, etc. virial coefficients obtained from the virial expansion as derivatives of the free energy cluster contributions in the virial formula given above \cite{EbFeKrRG19,KrEbCz75}.

The strong coupling contributions increase strongly with $e^2$
since $\psi_{ij} = O(e^4)$. As a result, the pressure may eventually become negative which would be unphysical.
Quite formally we may, in order to avoid negative values of $P_i$, transform the expression by replacing
$1 + g_i^{\text{sc}}$ which can get negative values by the strictly positive Pad\'e-like expression $1 / (1 - g_i^{\text{sc}})$
which would correspond to a geometric series and leads to
\begin{align}
 g_i = \frac{1}{(1 - g_i^{sc2} - g_i^{sc3} - g^{sc4} - ...)} - \frac{1}{6} z_i^2 \kappa \ell \cdot \sum_j \zeta_j^0 \cdot
 [{\tilde G}_0 + {\tilde G}_1](\eta_{ij}, \xi_{ij}) + g_i^{hc}.
\end{align}
The first term is now interpreted as the bound part of the ions of species i, which means that in the present approximation
there appears, as a degree of ionization, the expression
\begin{align}
{\tilde \alpha}_i =\frac{1}{1 + {\tilde \delta}^{(2)} + {\tilde \delta}^{(3)}+ {\tilde \delta}^{(4)} + ...}, \qquad {\tilde \delta}^{(k)} = (- g^{sck}) > 0 .
\end{align}
The ${\tilde \delta}^{(k)}$ are interpreted as degrees of association of $k$-clusters.
This leads finally to a rational virial representation of the osmotic pressure
\begin{align}
\frac{P_i}{(n_i k_{\text{B}} T)} = \frac{1}{(1 + {\tilde \delta}^{(2)} + {\tilde \delta}^{(3)} + {\tilde \delta}^{(4)} + ...)} - \frac{1}{6} z_i^2 \kappa \ell \cdot \sum_j \zeta_j^0 \cdot [{\tilde G}_0+ {\tilde G}_1] + g_i^{hc}.
\end{align}
This way without changing the accuracy in the linear order in density $O(n)$, we arrive at an mathematically equivalent but in the
context of association theory quite useful rational expressions for the pressure. We note that we may construct more powerful
Pad\'e-like rational expressions by including the higher association coefficients in nominator and denominator \cite{EbFoFi17}.
Finally, we note that the terms in the denominator may be interpreted as the effective degrees of association,
\begin{align}
{\tilde \delta}_i^{(2)} = \frac{1}{6} z_i^2 \kappa \ell \cdot \sum_j \zeta_j^0 \cdot [{\tilde G}_2(\xi_{ij}, \eta_{ij})],
\qquad {\tilde \delta}_i^{(3)} = \frac{1}{6} z_i^2 \kappa \ell \cdot \sum_j \zeta_j^0 \cdot [{\tilde G}_3(\xi_{ij}, \eta_{ij})].
\label{alphadelta}
\end{align}
Here, we included all contributions of higher order (not only their asymptotic)
into the expression for the degree of association, which makes some difference to the results of the previous section.
Consequently, the expressions for
$\alpha_i$ and the new here ${\tilde \alpha}_i$ as well as the old ${\tilde \delta}_i^{(k)}$ and the new ${\tilde \delta}_i^{(k)}$
are not identical, although they are in effect very close and converge to each other at small densities, the differences being
within the ``Onsager freedom in a ledger''. We remember the statement that chemical species are not uniquely defined, but depend
weakly on the context, i.e., on the imbedding into a theory \cite{Schroer11}.
The expressions for ${\tilde \delta}_i$  have a quasi-chemical meaning, since they may be interpreted as degrees of association. Formally, these new expressions may be obtained by a nonlinear extension of the canonical ensemble expansions assuming that these terms are part of a geometric series. The correctness of such an extension is justified by the studies within the grand-canonical ensemble \cite{FriedmanEb79,EbFoFi17}.

The terms ${\tilde G}_2$, ${\tilde G}_3$ stemming from second, third virial terms are here included into the strong coupling $g_i^{\text{sc}}$ since they describe the association effects
and give in our new approach the ``degree of association'' of pair formation
\begin{align}
 {\tilde \delta}_i^{(2)} = 8 \piup \sum_{j \ne i} n_j R_{ij}^3 \cdot (1 + \eta/3) \left(\frac{(\xi/(1 + \eta))^4}{24 (1 + 3 \eta)} +
\frac{(\xi / (1+\eta))^5}{120 (2 + 4 \eta)} + \frac{(\xi / (1 + \eta))^6}{720 (3 + 5 \eta)}+.. \right)
\label{deltafull}
\end{align}
Here, the circle is closing  since in the lowest approximation restricting to  $\xi^4$-term, we get a result close to
the previous result for a weak binding:
\begin{align}
{\tilde \delta}_i^{(2)} = 8 \piup \sum_{j \ne i} n_j R_{ij}^3 \cdot \frac{(1 + \eta /3 )}{(1 + 3 \eta)(1 + \eta)^4} \frac{\xi^4}{24}.
\label{alphaii}
\end{align}

\begin{figure}[!t]
\begin{center}
\includegraphics[height=6cm,width=5cm,angle=0]{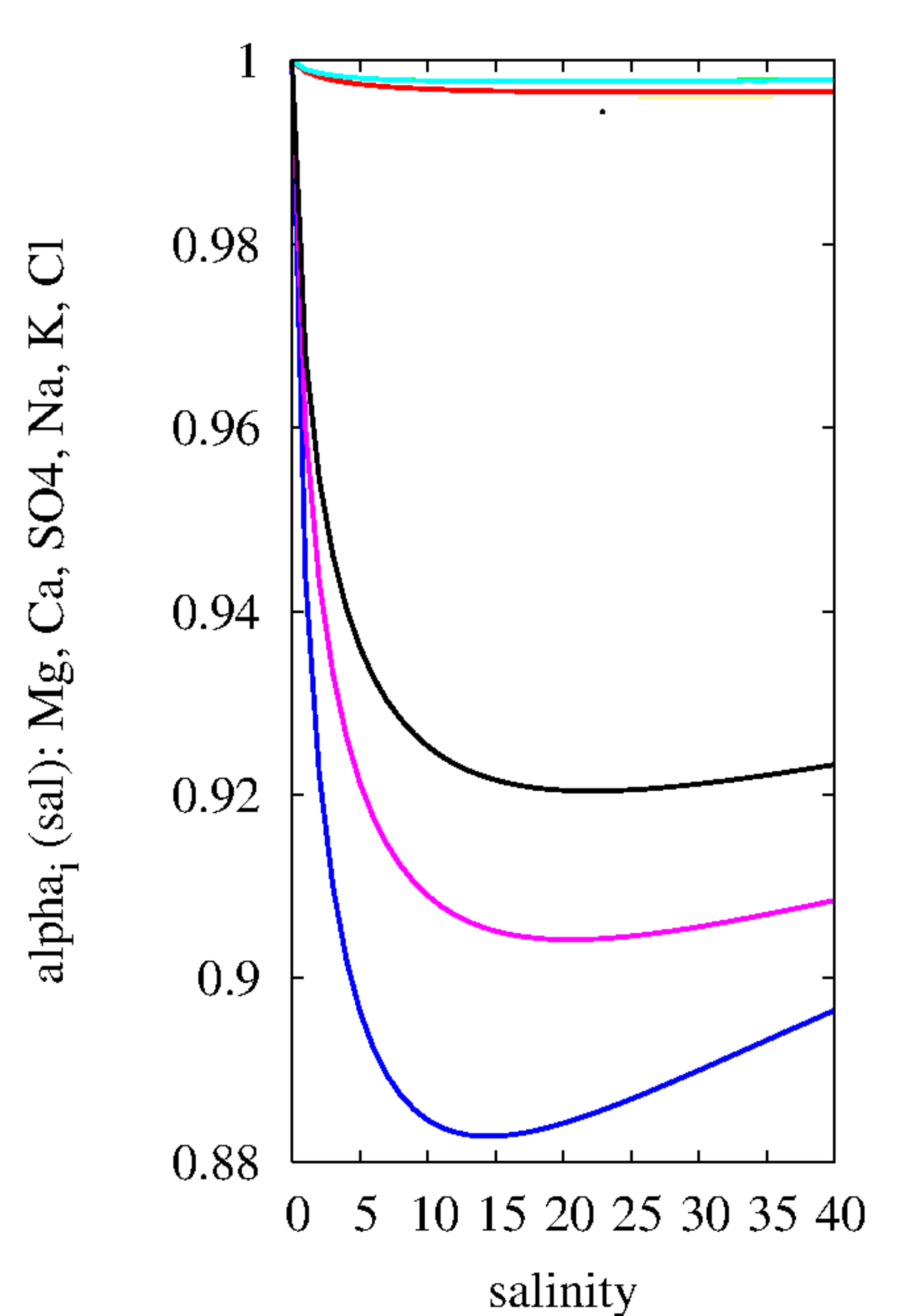}
\includegraphics[height=6cm,width=5cm,angle=0]{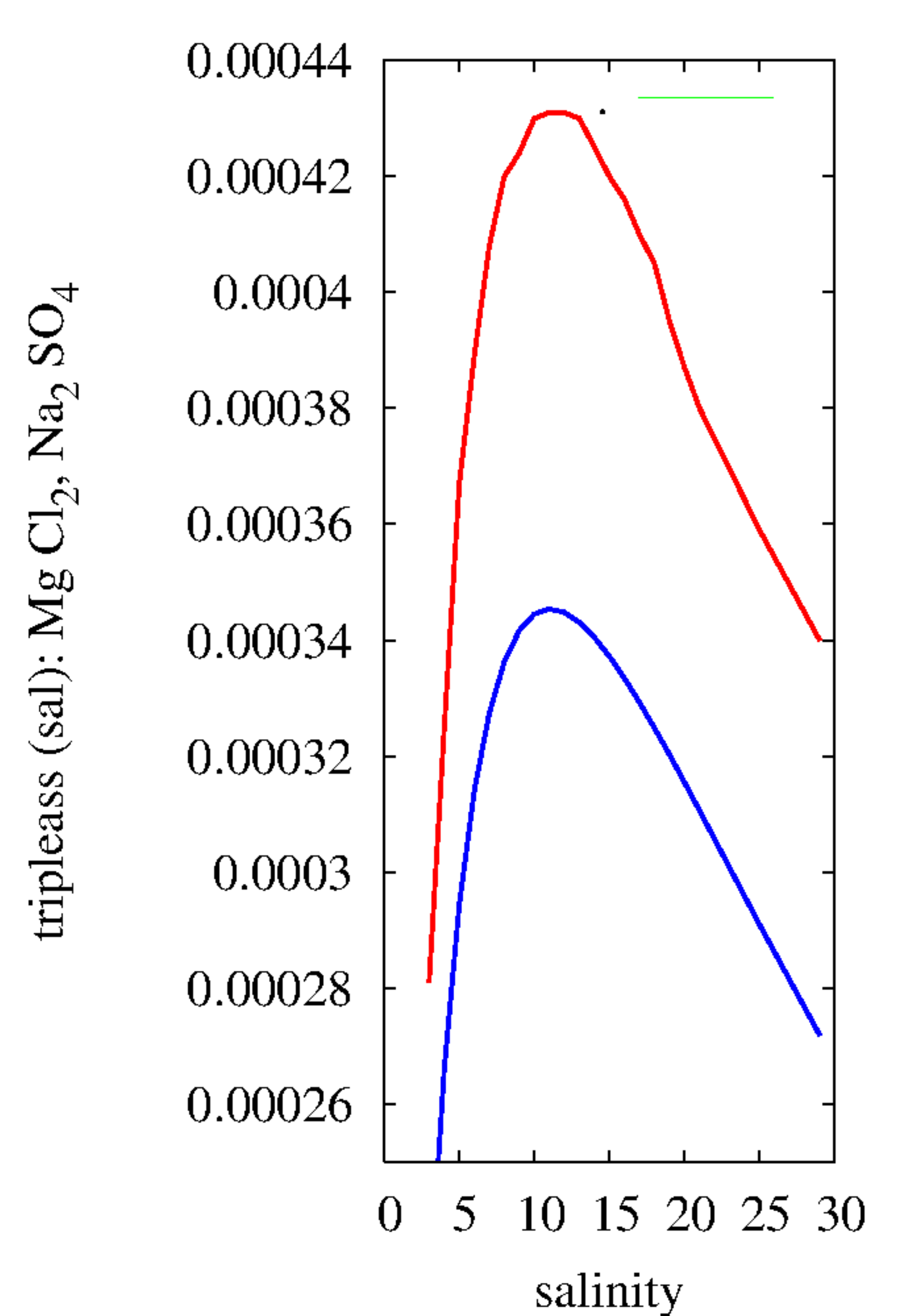}
  \caption{(Colour online) Seawater: Left-hand panel, degrees of ionization of ions as a function of salinity.
The curves describe (looking at salinity around the minima from below): Mg$^{2+}$, SO$_4^{2-}$, Ca$^{2+}$, Na$^+$,
 K$^+$, Cl$^-$. Right-hand panel, degree of triple association
 for the triples Mg Cl$_2$ (red) and Na$_2$ SO$_4$ (blue) depending on salinity.}
\label{IndSeai3}
\end{center}
\end{figure}

For small densities, we find by summing up the series in $\xi$, an expression valid for both versions of $\delta_i^{(2)}$-definitions for the pair formation which provides the degree of low-density ionization:
\begin{align}
\alpha_i = {\tilde \alpha}_i = \frac{1}{[1 + 2 \piup \sum_{j \ne i} n_j \ell_{ij}^3 \cdot n_2 (\xi_{ij})] }; \quad n_2 (x) = \sum_{k = 1,3,5, ..} \frac{x^k}{k (k + 3)!}.
\end{align}
This means that we have at low density a full agreement with the results of the previous section. We remember that $n_2(x)$ was previously discussed and expressed by $m(x)$ and the Kirkwood function. An application of the expressions
(\ref{deltafull}) and (\ref{alphaii}) to a
6-component model of seawater is given in figure~\ref{IndSeai3}.

The pressure approach leads to alternative expressions describing a weak pair association. It does not include assumptions about symmetries of the charges and may be applied to any neutral mixture of charges. For symmetrical systems, the odd terms cancel out and we get back familiar expressions used in many papers \cite{FaEb71,EbGr80}. The theory presented so far does not include  triple and higher association.
For MgSO$_4$ in standard seawater, the degree of association was measured based on the attenuation of sound by Fisher \cite{Fisher67}.
Based on these data, Fisher concluded that about 9.2 percent of the total Mg in seawater exists as MgSO$_4$. Kester and Pytkowicz estimated the mass action constants and for found for Mg, Ca a 
degree of association of about 10 percent \cite{KesterPytko69}. Our calculation gives for normal
ocean salinity for Mg a degree of association of about 12 percent, for SO$_4$ about 9 percent and for Ca  about 4 percent. Looking at the uncertainties of the experiment and the theory, the agreement seems to be satisfactory.

Now, we  propose  a similar procedure for calculating the degree of association of triples ${\tilde \delta}_i^{(3)}$.
The easiest way to find an estimate for the 3-particle association function is the effective charge method.
For the special case of a $(+)(2-)(+)$ or  $(-)(2+)(-)$ triple association, we find the degree of association of triples just
as a product two - particle terms modifying the charge. Further corrections arise by introducing some factors for the
interaction of equal charges. For example, for triples formed by an ion $i$ is with some higher charge $Z_i$ and the effective charge ${\tilde Z}_i $. In the lowest weak binding approximation, which includes only the terms of order $\xi^{10}$:
\begin{align}
{\tilde \delta}_i^{(3)} = 2 \piup^2 \ell^2 \sum_{j \ne i} n_j R_{ij}^2  \frac{\xi_{ij}^4}{24 (1 + \eta_{ij})^4 } \frac{(1 + \eta_{ij}/3)}{(1 + 3 \eta_{ij})} \cdot \sum_{k \ne i} n_k R_{ik}^2 \cdot \frac{\xi_{ik}^4}{24 (1 + \eta_{ik} )^4}
\frac{(1 + \eta_{ik}/3) }{(1 + 3 \eta_{ik})}.
\label{delta3}
\end{align}
We note that in the order $\xi^{10}$ this new pressure approach is compatible with equation~(\ref{eq27}).
According to this estimate, the triple association is as a rule, e.g., for seawater electrolytes, only a small correction, and the pair association dominates. A first application of our model to the triple association of ions in sea water was presented also in figure~\ref{IndSeai3} (right-hand panel).
We estimated the degree of triple association in seawater
for the triples MgCl$_2$ and Na$_2$SO$_4$. We find, according to out estimates,
that the formation of triples in seawater is rather seldom.
The predicted degrees of association to MgCl$_2$ and Na$_2$SO$_4$ obtained from our estimate equation~(\ref{delta3})
are in the range of $10^{-4}$ to $10^{-3}$. Qualitatively, the shape seems to be reasonable.
The maxima or minima, respectively, located as a rule near salinities between 10 and 15, demonstrate the typical screening effects. 
A stronger screening destroys Coulombic binding.

In the next section we are going from our extended physical approach as given by equations~(\ref{deltafull}--\ref{delta3}) where association constants do not appear in explicit way, to an alternative semi-chemical picture
where association constants $k_{ij...}$-s are used explicitly.

Summarizing, we developed here an ``extended physical description'' of electrostatic association based on rational expressions including higher cluster integrals. This way we also include "binding effects" in some order and find agreement with the semi-chemical formulae found in the last section. Our recipe is in brief as follows:\\
(i) Identify in the usual virial expansion of the pressure the ``binding contributions'' of plus-minus interaction which
provide big negative and for big $b_0$ asymptotically dominant effects. As we have shown such contributions are contributed
by terms of order $\xi_{+-}^4$ for pairing and by $\xi_{--}^2 \xi_{+-}^8$ or $\xi_{++}^2 \xi_{+-}^8$  for corresponding
triple formation.\\
(ii) Bring the ``binding contributions'' to the denominator anticipating that the corresponding geometric series stem
from the grand-canonical ensemble \cite{KeEb70,FriedmanEb79,EbFoFi17}.

We mention that the relation between the representations in the canonical and in the grand-canonical ensemble, which is not discussed here
in detail, is one of the keys for a deeper understanding of binding and non-binding contributions in cluster expansions. The idea is that
strong attracting contributions are better represented in the grand ensemble and repulsive contributions are better
represented in the canonical ensemble. The term ``better'' means here, that the convergence of the series is improved
\cite{EbFoFi17}.

Here, we were able to treat association without defining any mass action laws and mass action constants. Further, the present method predicts degrees of ionization $\alpha_i$ and degrees of association for pairs ${\tilde \delta}_i^{(2)}$ and triples, etc., for each of the ions. This allows us to describe a variety of electrolytes with multi-charged ions including models of seawater. A comparison of our theory for the corresponding activities with MC calculations by Ulfsbo et al. \cite{Ulfsbo15}
shows a reasonable agreement.
As a comment, we mention that the meaning of the word ``extended physical description'' is here that we specify in a physical description the main higher order
contributions which are responsible for binding effects. Our idea is that a complete physical description includes everything
and in particular also chemical effects. However, the problem is in this view, to identify and work out the
contributions relevant for association or chemical effects. We have described here this approach in some detail for pairing effects.
An extension to triple formation cannot be given at the same level of rigor due to the lack of detailed studies of the 3rd virial coefficient.
Our comparison with the existing data provided some hints to the existence of triples. We found that triples may be responsible for some corrections up to a few percent.

This new approach allows us to go further in improving a semi-chemical description and
to formulate the mass action laws. We will not go here the full way up to a complete chemical description. Instead, we go the half way to a more simple semi-chemical description
which works with simplified mass action laws and is comparable or even equivalent to our extended physical description. Our simplified semi-chemical approach is an approximation which works for
the case that association effects are weak. The treatment of strong association in a fully chemical approach without
simplification of the mass action law was discussed in \cite{KrBaHo00,EbGr80}.

\section{Semi-chemical treatment of pair and triple association}

Here, we show that the extended pressure approach suggested above is indeed half-way to a chemical description.
We study again only the model of charged hard spheres and follow the concepts developed previously in
\cite{FaEb71,Falkenhagen,EbKe66,FriedmanEb79,EbGr80,Justice,Justice1, Justice2}.

\begin{figure}[!t]
\begin{center}
\includegraphics[height=4.5cm,width = 9cm,angle=0]{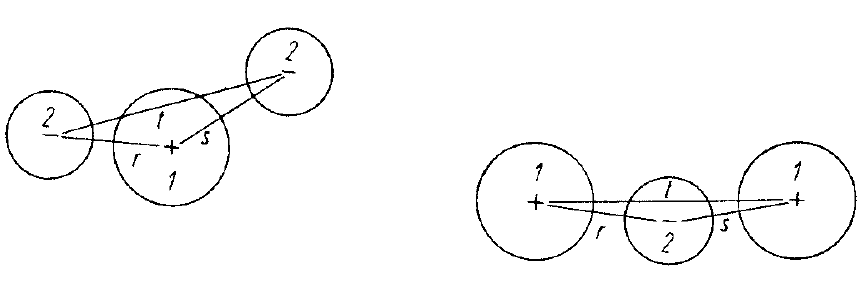}
\caption{Energy-rich configurations of ionic triples.
}
\label{3charges}
\end{center}
\end{figure}

The main effect of pair and triple association is the decrease of effective ion numbers from $n_i$ to $\alpha_i n_i$ where $\alpha_i$ is the degree of ionization. The following associates are of the main interest
\begin{align}
(+-) \quad (-+) \quad (+-+) \quad (-+-) \quad (+-+) \quad (-(2+)-) \quad (+(2-)+).
\end{align}
Examples of configurations of 3 ions in the plane which show high Coulombic energies
are shown in figure~\ref{3charges}.
As some interesting examples we look at the possible Coulombic associates of Cl$^-$ ions with Mg$^{2+}$  ions and with
La$^{3+}$ ions (see figure~\ref{3charges}).

The standard definition of a mass action constant applies to charge-symmetrical ionic associates
including, e.g., Mg$^{2+}$--SO$_4^{2-}$. More
difficult is the question how to define the association constant for triple associates as $(+)(--)(+)$ and $(-)(++)(-)$.
First we study  the maximal binding energy of 3 ions in linear order, e.g Cl-Na-Cl, Cl-Mg-Cl or Na-SO$_4$-Na possibly
one of them being double charged. The biggest energy has the
linear configuration $(-) (++) (-)$ or $(+) (--) (+)$ in the linear arrangement of the ions in contact. A simple estimate
of the energy of 3 ions in contact gives us
\begin{align}
E_{-+-} = E_{+-+} = (2 - 1/2 ) U_{0} = 1.5 U_0;\\
E_{-(2+)-} = E_{+(2-)+} = (4  - 1/2 ) U_0 = 3.5 U_0; \qquad U_0 = \ell / a.
\end{align}
These estimates show that we should not expect the formation of triples in the case of univalent ions,
since the formation of two separate pairs gives a lower energy than the formation of one triple.
However, in the case of divalent ions, the formation of triples is of advantage.
For the mass action constant of such triples, we expect the asymptotic
\begin{align}
k_{-(2+)-} = k_{+ (2-) +} \sim \exp[7 U_0 / 2 k_{\text{B}} T].
\end{align}
These results correspond to early estimates proposed by Kelbg,
Friedman and these authors \cite{EbKe66,KeEbKr68,FriedmanEb79} by using the results of mathematical studies
of cluster integrals \cite{EbKe66,Friedman,KrEbCz75,KrienkeSt84}.
This estimate was derived by using the assumption that the integrands have a sharp cusp at ion contact, which provides the asymptotic. This is an approximate approach to triple ionization which gives at least the correct asymptotic.

We consider now the configurations of 4 ions like in MgCl$_2$ solutions. In the case of quadruple formation,
we estimate an amount of 8 negative energy units stemming from 4 attractive Mg$^{2+}$-Cl$^-$ interactions.
Positive contributions come from the repulsion of the equally charged ions on opposite edges
which are approximately at a distance $\sqrt{5} R_{+-}$.
This estimate shows that these 4 ions would get in a solution more energy namely $4 E_0$ by forming two pairs instead of a quadruple due to the loss by positive repulsive contributions.
A first estimate of the energy of a LaCl$_3$-quadruple with La in the center and 3 Cl at the edges gives
\begin{align}
E_{\text{LaCl}_3} \simeq (9 - \sqrt{3}) U_0 \simeq 6.3 U_0.
\end{align}
Evidently, the formation of this triple is of some advantage in comparison to forming a pair with energy $3 U_0$ or a triple
with energy $6 - (1/2) U_0 = 5.5 U_0$ .

For a transition from a chemical to a semi-chemical approach, we  first discuss the pair association. In the standard approach
\cite{Falkenhagen,FaEb71,EbFeKrRG19,EbGr80}, the degree of ionization $\alpha$ is given by
the classical mass action law which we write in a form which is appropriate for iterative solutions
\begin{align}
\alpha = \frac{1}{1 + \alpha c (f_{\pm} (\alpha))^2 K(T) }.
\end{align}
The semi-chemical approach works with the first iteration following the zeroth approximation $\alpha^{(0)} =1$
\begin{align}
\alpha^{(1)} = \frac{1}{1 + c (f_{\pm} (1))^2   K(T) }.
\end{align}
The range of validity ends if more than $1/4$ of the ions are associated (see figure{\ref{Valleau}}). This is a strict assumption which still leads to a great simplification of the mathematics and is justified for many interesting systems as, e.g., seawater.
For the activity coefficients $f_{\pm}$  which appear in the mass action law, we use the
standard expressions for the electrical parts
\begin{align}
\ln f_{\pm} = \ln f_{\pm}^{el} = - \frac{z_+ z_-}{2 }\frac{\kappa \ell}{(1 + \gamma_{\pm})}.
\end{align}
We use here the so-called opposite-charge approximation (OPA), which is a specific property of Coulombic systems, based on the fact that in the region of stronger interactions (larger Bjerrum parameters), the encounter of opposite charges dominates
\cite{EbFoFi17}.
The main effect of pair and triple association is the decrease of effective particle numbers which leads to a decrease of the osmotic coefficients and the conductivities. We solve the MAL given above by iteration beginning with $\alpha = 1$. Our first and linear approximation, which works only for the regions where $\alpha $ is close to one, i.e., we are close to a full ionization,
is equivalent to the expressions in our semi-physical approximation. This interesting result means that physical and chemical expressions meet
here at half way.
In  case the association includes more than two ions, the MAL for the degrees of ionization looks as follows:
\begin{align}
\alpha_i =  \frac{1}{1 + \alpha_i \sum_{j \ne i} \alpha_j a_j (\alpha_j)  k_{ij}^{(2)} (T) + 2 \alpha_i \sum_{j,k \ne i, k}
a_j (\alpha_j) a_k (\alpha_k) k_{ijk}^{(3)} (T)}.
\label{alphabyit}
\end{align}
Again, the first order solution may be obtained by iteration starting with $\alpha_j = 1$ in the denominator.
This way, by extending our treatment of pair formation to higher association in first semi-chemical approximation,  we get the general expression
\begin{align}
\alpha_i =  \frac{1}{1 + \sum_{j \ne i} a_j k_{ij}^{(2)} (T) + 2 \sum_{j,k \ne i, k} a_j a_k k_{ijk}^{(3)} (T)},
\label{alpha0}
\end{align}
where the $a_i$ are the activities in electrical approximation:
\begin{align}
a_k = n_k f_k^{\text{el}}.
\end{align}

The definition of a mass action constant introduced above, applies to all charge-symmetrical ionic associates
including, e.g., Mg$^{2+}$-SO$_4^{2-}$. We have shown above that the association constant of pair formation starts with the order $e^8$
and for triple formation with $e^{20}$.
The restriction to these lowest orders gives the so-called weak binding constants of one multiple charged ions $Z >1$ with univalent ions
\begin{align}
k_{ij}^{(2)} (T)  = \frac{\piup}{3} \ell^3 Z^3 \xi_{+-}; \qquad k_{i,jk}^{(3)} (T) = \frac{\piup^2}{24^2} Z^4 \ell^6 \xi_{+-}^4\,.
\end{align}
These results correspond to the findings within our pressure approach in the previous section. We see that both approaches
agree in the lowest approximation. Introducing a new association function $n(x) = m(x) / x^3$, we may write down the results from the second section for triples in constant factor approximation
\begin{align}
k_j^{(2)} (T) = 2 \piup \ell_{ij}^3 n_2 (\xi_{ij}), \qquad n(\xi_{ij}) = m(\xi_{ij}) / \xi_{ij}^3, \qquad\\
k_{j,k}^{(3)} (T) = 8  \piup^2  [\ell_{ij}^3  n_2 (\xi_{ij})] [\ell_{ik}^3  n_2 (\xi_{jk}] [\exp(\xi_{jk}/2) - 1 + (\xi_{jk}/2] .
\end{align}
We study now some properties of the association functions useful for practical calculations \cite{EbGrRG22}.
\begin{align}
&n_2(b) \simeq \frac{b}{24}\cdot \left(1 + \frac{b^2}{90} + \frac{b^4}{8400} + \frac{b^6}{530000} \right) \quad \text{if} \quad b < 11,\\
&n_2(b) \simeq \frac{b}{24} \cdot \exp\left(\frac{b^2}{54}\right) \quad\quad\quad\quad\quad\quad\quad\quad\,\,\, \text{if} \quad b >11.
\end{align}
The mass action function $n_2(\xi)$ which determines the connection between the interaction parameters $\xi_{ij}$ and the association constants for for pair and triple formation
increases first nearly linearly in $\xi_{+-}$ then for $\xi_{+-} < 11 $ like the given polynomial
and for $\xi_{+-} > 11 $ the mass action constant starts growing exponentially.
\begin{figure}[!t]
\begin{center}
\includegraphics[height=5cm,width = 8cm,angle=0]{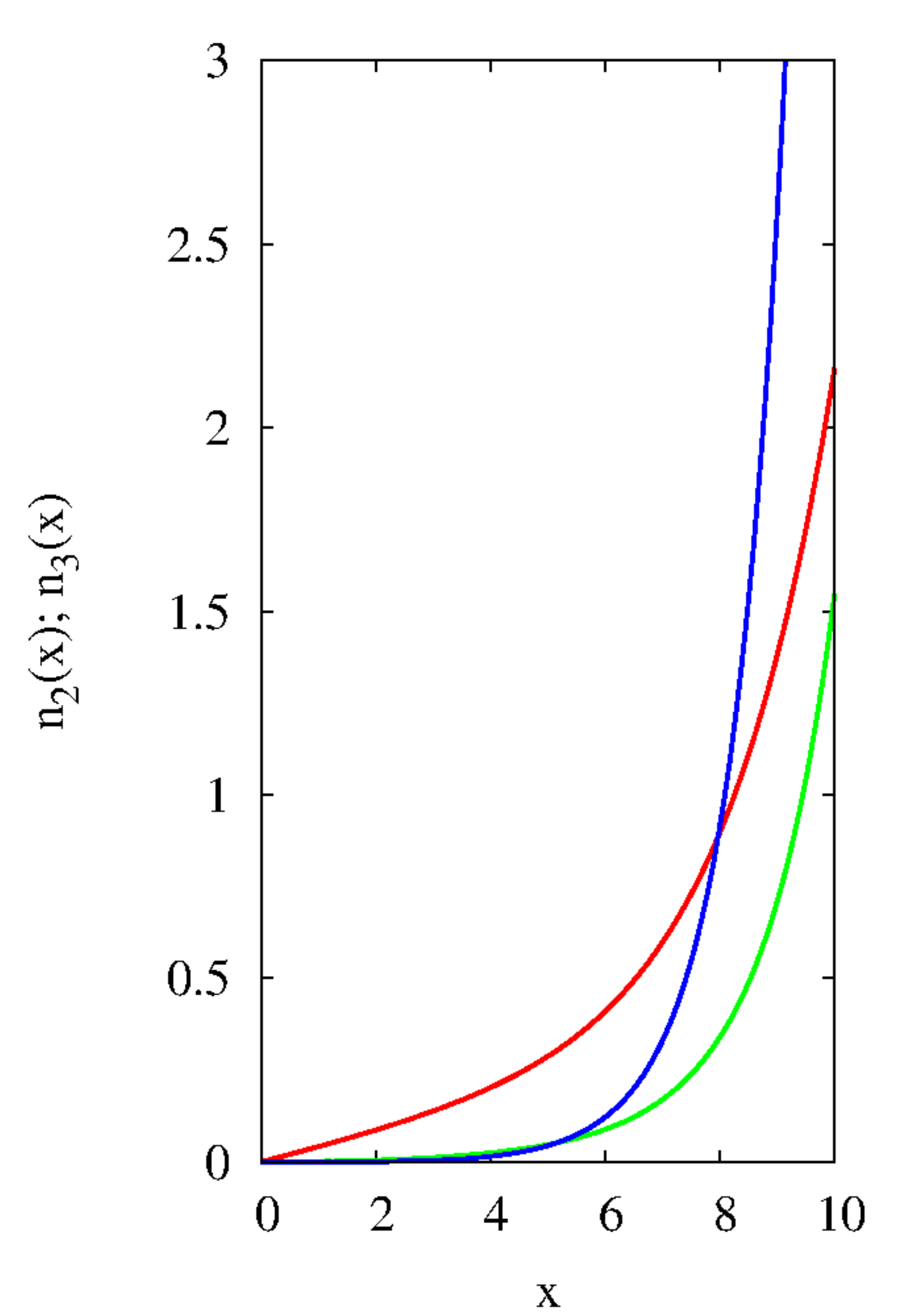}
  \caption{(Colour online) The pair association functions $n_2 (x)$ and the triple function $n_3(x)$ in dependence on the largest
Bjerrum parameter for the $(+-)$ interaction. For $n_2(x)$ we use an approximation by a polynomial valid for $x < 11$ (in red).
The lower curves show two approximations of the triple association functions, the effective charge approximation (in green) and
the constant factor approximation (in blue).
}
\label{MAC22}
\end{center}
\end{figure}
We may define an  $n_3$- function which depends on the ion parameters on 4 parameters in a complicated way
\begin{align}
k_3 (T) = 8 \piup^2 \ell_{+-}^6 \cdot n_3 (\ell_{+-}, \ell_{++}, \ell_{--}, R_{+-}).
\end{align}
A graphical representation of the triple association functions  depending on the plus-minus Bjerrum parameter is shown in figure~\ref{MAC22}. Note that we need two functions for triple association in the symmetric and in the asymmetric case for different combinations of the charges. For the relevant asymmetric charge combination with and without an approximation for the factor of repulsion, leads to the expression
\begin{align}
n_{3} = [\exp(x/2) - 1 + (x/2)] n_2(x)^2 \simeq (x^2 /8) n_2(x)^2.
\end{align}
A graphical representation of the functions $n_2(x), n_3(x)$ is given in figure~\ref{MAC22}.
Since a symmetrical combination of 3 charges
is energetically not favorable we concentrate here on the asymmetrical configurations ($+$) ($2-$) ($+$) and ($-$) ($2+$) ($-$).
Figure~\ref{MAC22} shows that for triple ionization, the constant factor approximation and the effective charge approximation approximately agree for $x < 6$; this is the region where triple association is relevant
for our applications, e.g., to K$_2$SO$_4$, Na$_2$SO$_4$ and MgCl$_2$, CaCl$_2$. In what follows we combine both methods which leads to the curves in between.

\section{Discussion of associating electrolytes with applications to K$_2$SO$_4$, CaCl$_2$ and LaCl$_3$}

\begin{table}[!b]
	\caption{Table of contact distances for several ion pairs including alkaline earth metal ions, sulfate ions and adapted ``ideal'' seawater ions according to \cite{EbFeKrRG19} with a new value for $R_{\text{CaCl}} = 500$.
		In the last but one column we give our new estimates for the ionization constants of several electrostatic pairs and triples in water at 25\textcelsius, The values for $n_3$ and $K_3$ are corrected by a factor due to the effect of charge asymmetry.
	}\vspace{.4cm}\centering
	\begin{tabular}{lcccccc}
		$a$-$b$-ions & $R_{ab}$     & $R_{aa}$       & $R_{bb}$   & $n_2; (n_3)$   & $K_2; (K_3) $  \\\hline
		Na-Cl       &  350           & 470   &  360  &  0.089    & $1.191 [\text{liter/mol}]$  \\
		K-Cl        &  320           & 400       &  360  & 0.099  & $0.548 [\text{liter/mol}]$  \\
		Mg-Cl     &  400           & 280       &  360   & 0.174  &  $7.71 [\text{liter/mol}]$ \\
		Cl-Mg-Cl     &  400           & 280       &  360  & 0.065 (0.204)  & $15.78 (50.26) [\text{liter/mol}]^2$\\
		Ca-Cl    &  500           & 320       &  360   &  0.131  & $5.81 [\text{liter/mol}]$  \\
		Ca-Cl$_2$     &  500           & 320       &  360 & 0.079   & $6.01 (19.45) [\text{liter/mol}]^2$ \\
		Na-SO$_4$   & 350           & 470       &  300   & 0.203  & $9.22 [\text{liter/mol}] $\\
		Na-SO$_4$-Na   & 350           & 470       &  300   & 0.381 & $30.40 (93.59) [\text{liter/mol}]^2$   \\
		K-SO$_4$    &  340          & 400       &  300   &  0.216   &  $9.57 [\text{liter/mol}]$  \\
		K-SO$_4$-K    &  340          & 400       &  300  & 0.432 &  $34.48 (106.1) [\text{liter/mol}]^2$ \\
		Mg-SO$_4$     &  290           &400      &  300   & 2.036  &  $721.2 [\text{liter/mol}]$  \\
		Ca-SO$_4$     &  350           & 380     &  380   & 1.219  &  $431.6 [\text{liter/mol}]$ \\
		La-Cl       &  270          &430       &  360   &  0.131  & $5.81 [\text{liter/mol}]$  \\
		La-Cl$_2$       &  270          &430       &  360  & 0.079   & $6.01 (19.45) [\text{liter/mol}]^2$\\
		Cd-Cl       &  420          & 300       &   360   \\
		Na-K        &  350          &  470         &  280  \\
		Na-Mg      &  300          &  470      & 380     \\
		Na-Ca       &  500          & 400       &  400     \\
		K-Mg        &  340          & 400         & 400     \\
		K-Ca        &  340         & 400         &320   \\
		Mg-Ca       &  330        &  440        & 440     \\
		Cl-SO$_4$     &  330        &  360        & 400   \\
		\label{TabContact}
	\end{tabular}
\end{table}

\begin{figure}[!t]
\begin{center}
\includegraphics[height=6cm,width = 9cm,angle=0]{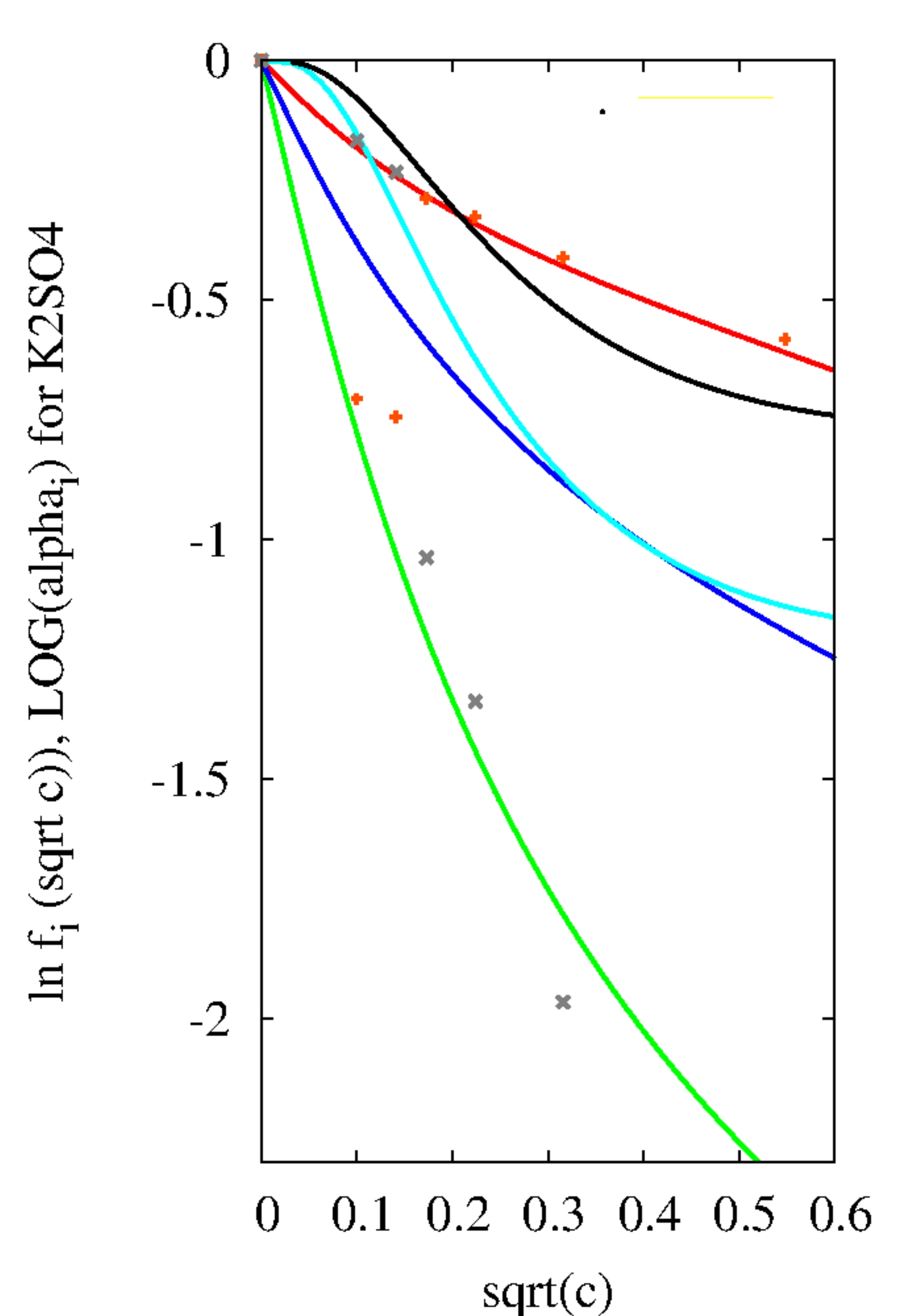}
\caption{(Colour online) Comparison of the activity coefficient of K$_2$SO$_4$ with data by Wilczek et al. denoted by points \cite{Vera}.
We show the activity of K$^+$ ions (in red) and below (in green) the activity
of SO$_4^{2-}$ in comparison with the data points by Wilczek et al. \cite{Vera}; the agreement is at least satifactory. The curve in between (in blue) is the mean activity of K$_2$SO$_4$.
The upper curves with turning points show the $\ln(\alpha_i)$ for the ions K$^+$ (in black) and SO$_4^{2-}$ (in turquoise).
}
\label{lnfKSO4}
\end{center}
\end{figure}

\begin{figure}[!t]
\begin{center}
\includegraphics[height=6cm,width = 9cm,angle=0]{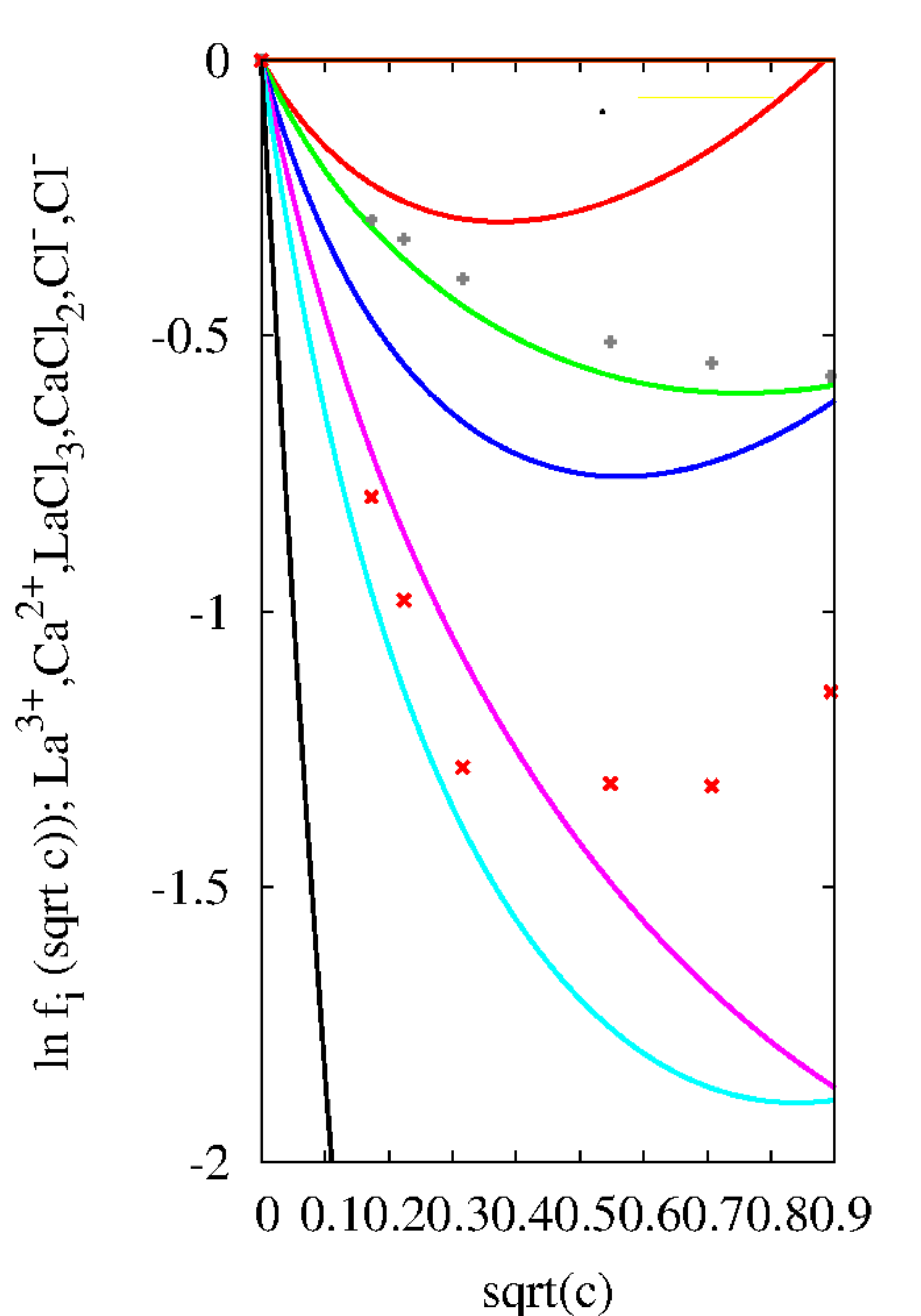}
\caption{(Colour online) We show the activity coefficients of several ions according to our theory, from above: Cl in CaCl$_2$ (red) and Cl in LaCl$_3$ (green), CaCl$_2$ (blue), LaCl$_3$ (magenta), Ca$^{2+}$ (turquise), La$^{3+}$ (black). We used here the extended physical approach neglecting quadruples, which may be the main reason for seemingly too low values for the activities of CaCl$_2$, La Cl$_3$, Ca$^{2+}$ and La$^{3+}$. The points denote the data measured by Wilczek-Vera et al.~\cite{Vera} for Cl$^-$ and La$^{3+}$ in LaCl$_3$.
The agreement with the measured data \cite{Vera} and with MDC calculations of Valisko and Boda for the mean activity coefficients is 
reasonable but not yet satisfactory \cite{ValiBoda}.
}
\label{lnfCLCaLa3}
\end{center}
\end{figure}

Here, we  treat the pair and triple association in aqueous solutions
based on the full nonlinear second and third virial coefficients. We may expect a corresponding contribution stemming from the strong coupling and negative definite part of the 4th virial coefficient.
Relevant for pair association are large negative parts from the 2nd virial coefficient, and important for triple association are the big negative
definite contributions from the 3rd virial coefficient. Following the general results from statistical thermodynamics
\cite{Falkenhagen,KrienkeSt84,EbKe66,Friedman,KrEbCz75,YukhnovskyHolovko80} we have shown that the key quantities for association are the asymptotically dominant parts of the strong
coupling terms in the cluster integrals.
Several examples of the resulting values of the association constants are given in table~\ref{TabContact}.
Note that in comparison to earlier work \cite{EbGrRG22} we introduced some corrections improving the dependence on the
$\ell_{++}, \ell_{+-}, \ell_{--}$ parameters. We compared here  two different ways of estimating the triple association constant, the constant factor approximation and the
effective charge approximation. Both methods are in reasonable agreement for $\xi_{+-} < 6$,  and then they start to disagree.
In the constant factor approximation, the triple association constants for salts like K$_2$SO$_4$, CaCl$_2$ and ions like (LaCl$_2)^+$ are estimated again expanding the repulsive factor which leads to the expressions
\begin{align}
k_{\text{KSO}_4\text{K}} &=  \piup^2 \ell^2 R_{\text{KSO}_4}^4 \cdot  m (\ell_{\text{KSO}_4}/ R_{\text{KSO}_4})^2, \nonumber \\
\noindent k_{\text{ClCaCl}} &=  \piup^2 \ell^2 R_{\text{CaCl}}^4 \cdot  m (\ell_{\text{CaCl}}/ R_{\text{CaCl}})^2, \qquad \nonumber \\
\noindent k_{\text{ClLa}+ \text{Cl}} &=  \piup^2 \ell^2 R_{\text{LaCl}}^4 \cdot  m (\ell_{\text{LaCl}}/ R_{\text{LaCl}})^2.
\end{align}
Several numerical values of mass action constants for pair and triple formation estimated this way are given in table~\ref{TabContact}.
A comparison of the predicted individual activities with measurement is difficult due to the lack of data. A preliminary
comparison of our results for the individual activity coefficients using the data of Wilczek et al. \cite{Vera} and of Valisko and Boda is shown in figures~\ref{lnfKSO4} and \ref{lnfCLCaLa3}. For the salt K$_2$SO$_4$, our prediction for the individual activities of
the ions K$^+$ and SO$_4^{2-}$ is at least close to the data found by Wilczek et al \cite{Vera}. For the salt CaCl$_2$, we may compare with the results by Wilczek et al. and Valisko and Boda \cite{ValiBoda}. Again, the agreement of our results with the data of other workers
for CaCl$_2$ is sufficient, although not quantitative.
For LaCl$_3$, our theory is not capable of reproducing the pronounced
minimum around $\sqrt{c} \sim 0.4 (\text{mole/liter})^{0.5}$ predicted in \cite{ValiBoda}. The reason is possibly that we did not include so far the fourth virial coefficient. We underline that our approach provides also the individual activities. 

For the fourth cluster integral and the corresponding association constants, a consequent statistical analysis is still missing.
Therefore, we restrict ourselves here to an estimate following the lines valid for the third virial coefficient.
In order to estimate the quadruple association constant for a salt like LaCl$_3$, we fix the Cl-Cl-distances at some energetically favorite distance like $\sqrt{3} R_\text{LaCl}$, which we guess from figure~\ref{IndSeai3}. Then, the factors for the repulsive terms may be taken out, which leads to factorization of the remaining terms in the integral and again we get approximating  the repulsive term by the quadratic order
\begin{align}
k_{\text{LaCl}_3} = C \piup^3 R_{\text{LaCl}}^3 \ell^6_{\text{ClCl}}\cdot [m (\ell_{\text{LaCl}}/ R_{\text{LaCl}})]^3.
\label{kLa3Cl}
\end{align}
We estimated several values of  the quadruple association constant following equation~(\ref{kLa3Cl}) and found now
$C \simeq 1/3456$.
These and other values for the association constant that we obtained this way are given in table~\ref{TabContact}. Some of the values were given already in \cite{EbFeKrRG19} and \cite{EbGrRG22}. We  added a few not so well studied ions as Cd and La. The crystallographic radius for Cd$^{2+}$ is with 95 pm just a few pm higher than that for Mg$^{2+}$ which is 86 pm. Therefore, we may assume that the contact distances in a solution are also close, we took $R_{\text{CdCl}} = 420$ pm. For La$^{3+}$, we know that the crystallographic radius is smaller than that for Mg$^{2+}$. Following canonical MC simulations by Valisko and Boda (2017, 2018),  we assume for those ions in water $R_{\text{LaLa}} = 430$ pm and $R_{\text{LaCl}} = 270$ pm corresponding to a quite large Bjerrum parameter $\xi_{\pm} = 7.95$.
Several applications  to the binary electrolytes CaCl$_2$ and LaCl$_3$ are presented in figure~\ref{lnfCLCaLa3}. So far, the agreement with available data is not yet quantitative \cite{ValiBoda,Vera}.
Note that we differ between association constants $k_{i,...}$ in the density scale of statistical mechanics particle $\text{number/cm}^3$, and $K_{i...}$ in
the usual chemical concentration scale mol/liter. In order to switch between the figures in both of the scales, we
remember the relations
\begin{align}
n_i [cm^{-3}] = 6.023 \cdot 10^{20} c_i [\text{mol/liter}]; \qquad K_2  = 6.022 \cdot 10^{-4} k_2.
\end{align}
Using these factors of recalculation we derived the numbers given in the last column of table~\ref{TabContact}.

\section{Conclusions}
In the present survey we discuss ion association and activity coefficients in 1-1, 1-2, and 2-2 electrolytes.
For 1-3-electrolytes, we restrict ourselves to some qualitative analysis. The association constants for triple and quadruple association are estimated in the effective charge and constant factor approximation. We repeat and correct several fully analytical
results to calculate the degrees of weak association and activity coefficients \cite{EbFeKrRG19,EbGrRG22}.
In particular, we discuss extended physical methods based on rational extensions of pressure virial expansions,
which do not use mass action laws in an explicit way. Further, we use semi-chemical methods which define the mass action constants for weak association but simplify the mass action laws. This way the restriction to weak association allows us to avoid the use of full (nonlinear) mass action laws. We estimate the ionization constants for pair and triple electrostatic association from the cluster integrals and calculate the degree of ionization for electrolytic mixtures including seawater in the regions of weak association,
in general  20--30 percent smaller.
Several applications  to the electrolytes CaCl$_2$ and LaCl$_3$ are presented in figure~\ref{lnfCLCaLa3}.

Summarizing our findings: based on the results of statistical physics, we recommend in addition to  standard methods of calculating the individual activities,
new statistical tools for the calculation of individual and mean activities and degrees of association of ions from lower concentrations up to moderate concentrations/salinity.
The methods are based on the model of hard spheres with non-additive radii in combination with the nonlinear Debye-H\"uckel (or mean spherical) approximations for screening.
Association effects are included by rational virial expressions for the osmotic pressure, which take into account higher order
terms stemming from the grand canonical ensemble. These methods avoid the solution of nonlinear mass action laws and use instead
rational polynomials which include terms from the fugacity expansions of the pressure.
For the first time, pair and triple association is taken into account on same footing in a systematic way; an extension to quadruple association is proposed. A table of most relevant association constants for of electrolytes including seawater ions is given. We use for all calculations only hard-charged sphere models
and, as parameters, the charges and non-additive contact distances. The proposed formulae are fully analytical and results can be obtained on a home computer. We stay within the traditional physical and chemical approaches and develop a rational virial
approach to the pressure and a semi-chemical description of associates. In the purely physical approach to the osmotic
pressure based on rational expressions avoiding explicit definitions of bound pairs or triples. In our semi-chemical approach, explicit association constants are defined and calculated; this is based on the concept that binding is due to higher powers in the interaction determining the asymptotic properties of the cluster coefficients. In both approaches, we concentrate on the relevant contributions of strong Coulombic interactions to association through the higher virial coefficients which determine association effects. Note that both approaches are in full agreement at low densities.
The input parameters needed in our theory are, beside dielectric constants, the contact distances of all ion pairs.

In conclusion: the present approach is based on the idea that bound states need a special statistical treatment, although here we have  some freedom. We provide for associating electrolytes as well as for
seawater a reasonable simple treatment and at least qualitative agreement with other available results. Our input data are  3 contact distances for binary electrolytes and about 20 adapted contact distances for standard seawater, which may depend on temperature and pressure.
So far, all results are given only for the temperature 25\textcelsius, an extension to other temperatures is easy,
we have only to change the numerical value for the Coulomb length $\ell$. Note that in other approaches to seawater, the number of free parameters is still larger. On the other hand, our basic parameters, i.e., the ionic contact distances have a clear physical interpretation.

\section*{Acknowledgement}
We express our thanks to many colleagues for suggestions and encouragement. As already mentioned, the methods we developed here originated in 
intensive discussions which we had long time ago with Professor G\"unter Kelbg in Rostock and Professor Ihor Yukhnovskii during his visits in the 60th and 70th to Rostock and our visits
in the 60th and 70th to Lviv \cite{KeEb70}. We feel that we are standing on the shoulders of these two pioneers and express our deep gratitude. 
For more recent discussions we thank our colleagues Rainer Feistel (Rostock), in particular for suggestions about the relations
to the tasks of seawater research and Manfred Grigo (Rostock) for an intensive collaboration and common works about the association theory \cite{EbFeKrRG19,EbGrRG22}.
Further, we express our thanks to Grazyna Wilczek (Montreal), for providing material about activity coefficients and tables with data.
Finally, we sincerely thank Myroslav Holovko (Lviv) for many fruitful discussions and substantial support of this work.

\ukrainianpart

\title[Статистична теорія парціальних коефіцієнтів активності електролів з врахуванням множинних іонних зарядів]
{Статистична теорія парціальних коефіцієнтів активності електролів з врахуванням множинних іонних зарядів
}
\author[В. Ебелінг, Г. Крінке]{В. Ебелінг\refaddr{label1},
	Г. Крінке\refaddr{label2}}

\addresses{
	\addr{label1} Інститут фізики, Університет ім. Гумбольдта, Берлін, Німеччина
	\addr{label2} Інститут фізичної хімії, Університет Регенсбурга, Німеччина}

\makeukrtitle

\begin{abstract}
	У попередній роботі ми розробили новий статистичний метод для розрахунку парціальних активностей іонів включно з іонною асоціацією. У даній роботі вивчаємо багаточастинкові електростатичні взаємодії, пов’язані з вищими кластерними інтегралами, та визначаємо константи іонізації в законі діючих мас для асоційованих іонних кластерів. На відміну від теорії Б’єррума та Фуосса, наша концепція асоціації базується не на просторових критеріях, а на інтенсивності взаємодії, яка задається розвиненнями за параметром Б’єррума ($e^2 / D_0 k_{\text{B}} T a$, де $a$--контактна відстань) і визначається асимптотичними властивостями кластерних інтегралів. При утворенні іонної пари наша константа в законі діючих мас є класичним аналогом відомої функції розподілу Планка для водню. Зазвичай, нові константи асоціації при певних значеннях параметрів взаємодії майже вдвічі менші, ніж величини, отримані на основі традиційних виразів (наприклад, Фуосса і Крауса). У роботі вивчається кілька іонних систем, зокрема, CaCl$_2$, MgCl$_2$, Na$_2$SO$_4$, K$_2$SO$_4$, LaCl$_3$ та модель морської води. Для кількох асоціативних електролітів і морської води досягнуто хорошого узгодження з експериментами та результатами моделювання Монте-Карло.
	\keywords статистична фізика, термодинаміка, електроліти, коефіцієнти активності, іонна асоціація, морська вода
\end{abstract}

\end{document}